\documentclass[twocolumn]{aastex63}

\usepackage{graphicx}
\usepackage{dcolumn}
\usepackage{bm}

\usepackage{makecell}
\usepackage{verbatim} 
\usepackage{amsmath}

\begin{document}


\title{Magnetic Drag and 3-D Effects in Theoretical High-Resolution Emission Spectra of Ultrahot Jupiters: the Case of  WASP-76b }

\correspondingauthor{Hayley Beltz}
 \email{hbeltz@umich.edu}

\author[0000-0002-6980-052X]{Hayley Beltz}

 \affiliation{Department of Astronomy, University of Michigan, Ann Arbor, MI 48109, USA}

 \author[0000-0003-3963-9672]{Emily Rauscher}
\affiliation{Department of Astronomy, University of Michigan, Ann Arbor, MI 48109, USA}

\author[0000-0002-1337-9051]{Eliza M.-R.\ Kempton}
\affil{Department of Astronomy, University of Maryland, College Park, MD 20742, USA} 
 
 \author[0000-0003-0217-3880]{Isaac Malsky}

 \affiliation{Department of Astronomy, University of Michigan, Ann Arbor, MI 48109, USA}

\author[0000-0003-1092-6520]{Grace Ochs}
\affiliation{Department of Astronomy, University of Michigan, Ann Arbor, MI 48109, USA}
\author[0000-0001-6268-5517]{Mireya Arora}
\affiliation{Department of Astronomy, University of Michigan, Ann Arbor, MI 48109, USA}
\author[ 0000-0002-2454-768X]{Arjun Savel}
\affil{Department of Astronomy, University of Maryland, College Park, MD 20742, USA}

\begin{abstract}
Ultrahot Jupiters are ideal candidates to explore with high-resolution emission spectra. Detailed theoretical studies are necessary to investigate the range of spectra we can expect to see from these objects throughout their orbit, because of the extreme temperature and chemical longitudinal gradients that exist across day and nightside regions. Using previously published 3D GCM models of WASP-76b with different treatments of magnetic drag, we post-process the 3D atmospheres to generate high-resolution emission spectra for two wavelength ranges and throughout the planet's orbit. We find that the high-resolution emission spectra vary strongly as a function of phase, at times showing emission features, absorption features, or both, which are a direct result of the 3D structure of the planet. At phases exhibiting both emission and absorption features, the Doppler shift differs in direction between the two spectral features, making them differentiable instead of canceling each other out. Through the use of cross-correlation, we find different patterns in net Doppler shift for models with different treatments of drag: the nightside spectra show opposite signs in their Doppler shift, while the dayside phases have a reversal in the trend of net shift with phase. Finally, we caution researchers from using a single spectral template throughout the planet's orbit; this can bias the corresponding net Doppler shift returned, as it can pick up on a bright region on the edge of the planet disk that is highly red- or blue-shifted. 
\end{abstract}

\section{Introduction}
High-Resolution Spectroscopy (HRS) ($R \gtrsim 15,000$) has opened up exciting new pathways into exoplanet atmospheres. By taking advantage of the tens of thousands of spectral lines separable at this higher resolution, astronomers are able to detect scores of atmospheric species, despite the orders of magnitude flux contrasts between the planet and the star. In order to unlock this information, the HRS method uses a template spectrum for the planet and cross-correlates this across wavelengths (which is equivalent to varying the Doppler velocity shift of the spectrum) with the data to search for a peak signal. The planet's orbital motion will cause its signal to shift with wavelength during an observation. Since the orbital motion of the planet is known, this method can reject spurious low-significance signals and reliably pinpoint the planet's atmospheric signature. The atmospheric conditions used to calculate the template spectrum can be adjusted until they most closely match reality, producing the largest detected signal. HRS can also be used to detect atmospheric motion on the planet in the form of net redshifts or blueshifts \citep[such as the first-ever HRS detection from ][]{Snellen2010} as well as constrain rotational velocities from Doppler broadening \citep{Snellen20104,Brogi2016,Schwarz2016,Flowers2019} and even obliquity \citep{Bryan2020}. Recently, the most precise C/O ratio in a hot Jupiter was measured with HRS techniques \citep{Line2021} using the framework described in \citet{Brogi2019}. An excellent review of HRS techniques can be found in \citet{Birbky2018,Brogibook2021}.

Crucial to the successful use of this method is having a template spectrum to use in the cross-correlation that accurately matches the planet's true spectrum.  For this work, we will focus on the two classes of model spectra: those generated from 1D atmospheric models and those generated from more physically consistent 3D models, each of which have their advantages and disadvantages. On one hand, 1D models are feasible to compute over large parameter spaces in a reasonable amount of computing time. They have also been used with high-resolution emission spectra to detect a variety of species including (but certainly not limited to) Fe, Ti, Mg, TiO, CO, OH, H$_{2}$O,CH$_{4}$  \citep[see][for some recent examples]{Brogi2014,Gandhi2018,Guilluy2019,Nugroho2020,Giacobbe2021,Kasper2021,Nugroho2021,Pelletier2021}. However, planets are 3D objects, and using a 1D template can lead to unintended consequences. At low and medium resolution, using a 1D model can bias interpretations such as incorrectly constraining abundances and temperature structures \citep[][]{Feng2016,Blecic2017,Caladas2019,Pluriel2020,Taylor2020}. These biases should become more pronounced when planetary properties vary strongly as a function of longitude. In HRS observations, spectral liens are more fully resolved, allowing the 3D atmospheric structure to directly impact the shapes of the observed lines, calling into question the applicability of the 1D assumption.  

3D models, on the other hand, are much more computationally expensive to calculate (generally requiring at least a few weeks to run to completion). 3D models are also by nature more complicated to calculate simulated spectra from as the geometry and radiative transfer calculations are more complex than the 1D case \citep{Zhang2017}. (However, recent open-source tools such as the ones presented in \citet{Lee2021} are making this barrier less steep.) Importantly, if the 3D model better matches reality, its spectrum will produce a stronger cross-correlation signal with the high-resolution data. In \citet{Beltz2021a}, we found that our 3D hot Jupiter model outpreformed hundreds of 1D models when used in the analysis of high-resolution emission spectra. 3D models are also able to capitalize on the 3D effects present in HRS such as Doppler shifting of lines, temperature gradients, and chemical inhomogenieties \citep{Flowers2019,Macdonald2021,Prinoth2021}. Other recent work has also been investigating other routes to retrieve 3D structure from high-resolution emission spectra \citep[][]{Herman2022,vansluijs2022}

3D models are especially necessary for ultrahot Jupiters (UHJs), given the large spatial variations across these planets due to their extreme irradiation regime. 
These planets, with equilibrium temperatures that exceed $\sim 2200$ K, orbit so close to their host star that they are expected to have tidally-synchronized rotation states \citep[][]{Showman2002}. This means that these planets have a constant dayside that is continually bombarded by stellar irradiation and a permanent nightside. An emerging ubiquitous property of UHJ atmospheres is that they exhibit dayside temperature inversions---that is, regions of decreasing temperature with increasing pressure---in contrast to their cooler hot Jupiter cousins \citep[][]{Baxter2020,mansfield2021}. The exact cause of these inversions is still debated, but high-altitude optical and UV absorbers such as TiO or VO \citep[][]{Hubney2003,Fortney2008} or Fe, SiO, and metal hydrides \citep[][]{Lothringer2018,Gibson2020,Lothringer2022} are likely candidates.  Alternative potential causes for the inversions include high C/O ratios \citep{Molliere2015} or inefficient IR cooling \citep{Gandhi2019}. These temperature inversions manifest themselves in the emitted spectra of the planet, where dayside spectra show emission features. The nightsides of these planets, due to their mostly monotonic temperature structure, will only show absorption features.

Due to the extreme irradiation received, the daysides of UHJs are  hot enough for the thermal dissociation of important molecules. This can significantly change the main opacity sources \citep[e.g., removing water and adding H$^{-}$, such that spectral features become more muted;][]{parmentier2018} or in the most extreme cases change the dominant atmospheric gas from molecular to atomic hydrogen \citep{Bell_2018}. The dissociation of molecular hydrogen on the dayside and its recombination on the nightside should decrease the temperature gradient across the planet and decrease the expected wind speeds \citep{Tan_2019}.

For this work, we focus our investigation on magnetism and its role in UHJ atmospheres. Magnetic field generation strength still holds uncertainties, but work from \citet{Christensen2009} predicts that given a sufficient rotation rate (like that of tidally locked UHJs) magnetic field strength of planets may scale with energy flux. For the most inflated HJs, \citet{Yadav2017} estimates maximum surface field strengths to be between 50-100 G, with more massive planets having the potential for even stronger fields. While there have been no direct measurements of exoplanet magnetic field strength yet, several indirect methods can offer constraints for particular planets.  Recently, \citet{BenJaffel2021} detected the magnetosphere of the warm Neptune HAT-P-11b and estimate the equatorial magnetic field strength to be between 1--5 G. Magnetic planet star interactions, specifically Calcium II K emission, have also been used to indirectly estimate surface magnetic field of hot Jupiters in the range of 10-100 G \citep[][]{Cauley2019}. Variability in hot spot location has been found in 3D non-ideal MHD models, leading \citet{Rogers2017} to estimate a minimum magnetic field strength of 6 Gauss for the UHJ HAP-P-7b, in order to explain observed variability on that planet \citep[it should be noted however that recent work from ][find that that the observed variation could be due to non-atmospheric sources]{lally2022}. In our previous work, \citet{Beltz2022}, we were able to estimate a minimum field strength of 3G for WASP-76b based on comparisons with phase curves published by \citet{May2021}.  

In our modeling, we remain agnostic to the intricacies of the planet's internal dynamo and instead  assume a deep-seated global magnetic field of some magnitude exists, in order to focus on the effects of magnetic drag in the atmosphere of the representative UHJ WASP-76b and how that drag influences its emission spectra. Notably, our magnetic drag treatment does not solve the full non-ideal MHD equations that, due to their added computational complexity, require other simplifying assumptions \citep[see the discussion of this in ][]{Beltz2022}. Instead, we use a parameterized approximation of magnetic drag first introduced in \citet{Perna2010magdrag}. A physical explanation of this parameterization can be thought of as follows: due to the partial thermal ionization of species on the dayside of UHJs, the strong winds on the planet will blow these charged particles (embedded in a mostly neutral flow) around the planet, crossing magnetic field lines and generating current. These currents, should they travel into the planet's interior and deposit heat in a process known as Ohmic dissipation, could play a role in inflated radii of many hot Jupiters \citep{BatyginStevenson2010,Perna2010ohmic,Thorngren2018}. In the atmosphere, the bulk Lorentz force that ionized particles experience can reduce circulation efficiencies and therefore increase the day-night contrast \citep[][]{Perna2010magdrag,Menou_2012,Batygin2013,Rogers_2014_showman}. Previous work showed this in models of the hot Jupiters HD209458~b and HD189733~b \citep[][]{RauscherMenou2013} and for the case of the ultrahot Jupiter WASP-76b \citep{Beltz2022}. Now, in this work, we take the models from \citet{Beltz2022} and post-process their results to study the resulting high resolution emission spectra. 

Our models are all based on parameters from the UHJ WASP-76b, an inflated planet orbiting its host star every 1.81 days. This planet has been the target of many high-resolution transit analyses, and a multitude of different species have been detected in its atmosphere, including Fe I, Na, Li I, Ca II \citep{Seidel2019,Deibert2021,Cassayas2021,Kawauchi2021,Taberno2021},
and recently molecules like  OH \citep{Landman2021}.Analyses of high resolution emission spectra have yet to be published.  The work by \citet{Kesseli2021} explores an entire atomic spectral survey of the atoms and ions that have been detected on this UHJ and finds evidence for temperature and chemical gradients on the planet. This planet has also garnered interest from astronomers due to an asymmetric detection of iron in its transmission spectrum throughout transit, first found by \citet{Ehrenreich2020} and later confirmed by \citet{Kesseli2021wasp76}. \citet{Ehrenreich2020} proposed this asymmetric absorption signal is a result of iron condensing on the nightside of the planet. Other proposed mechanisms for this absorption signal include temperature asymmetry between the trailing and leading limbs \citep{Wardenier2021} or clouds \citep{Savel2021}. 3D atmospheric models of this planet have also been published in \citet{May2021,Savel2021,Wardenier2021,Beltz2022}.

In this work, we generate high-resolution emission spectra from previously published models of the UHJ WASP-76b to explore how different magnetic drag treatments affect the resulting spectra and how these differences change as a function of phase. This particular planet is of high observational interest but also provides a route to make predictions regarding emission spectra from UHJs more generally.  The structure of this paper is as follows: In section \ref{sec: Methods}, we briefly describe the models used in this work, the variety of drag treatments in our models, and the process of post-processing our model output to create simulated high-resolution emission spectra. In section \ref{sec: Results}, we explore the features that show up in our 3D emission spectra and how they differ between our models. In section \ref{sec:Discussion} we discuss important caveats to our models and assumptions. Finally, we summarize our main findings in section \ref{sec:Conclusion}.

\section{Methods} \label{sec: Methods}
\subsection{General Circulation Model}
In this work, we use a subset of the models produced and described in \citet{Beltz2022}, and will briefly summarize the details here. General Circulation Models (GCMs) solve the primitive equations of meteorology in three dimensions to simulate the physics and circulation patterns of planetary atmospheres. We use the GCM from \citet{Rauscher2012GCM} which underwent a radiative transfer upgrade in \citet{newradRomanRausher} \citep[based on][]{Toon1989}. The absorption coefficients for the WASP-76b models were informed by the 1D temperature-pressure profiles modeled in \citet{Fu2021}, chosen such that the double-gray analytic profile \citep{Guillot2010} roughly matched those results. Every model was run with 65 vertical layers spread out evenly in log space over 7 orders of magnitude in pressure from 100 to $10^{-5}$ bar and with a horizontal spectral resolution of T31, corresponding to $\sim 4$ degrees at the equator. Each simulation was calculated for a total of 2000 planetary days. Relevant global parameters are shown in Table \ref{tab:gcm_params}.

\begin{deluxetable}{lc}
\caption{WASP-76b Model Parameters} 
\label{tab:gcm_params}
\tablehead{ \colhead{Parameter} & \colhead{Value}} 
\startdata
         Planet radius, $R_{p}$ & $1.31 \times 10^{8}$ m \\
         Gravitational acceleration, $g$ & 6.825 m s$^{-2}$ \\
         \added{Orbital Period} & \added{1.81 days} \\
        Rotation rate \tablenotemark{a}, $\omega_{\mathrm{orb}}$ & $4.018 \times 10^{-5}$ s$^{-1}$ \\
         Substellar irradiation, $F_{\mathrm{irr}}$ & $5.14 \times 10^{6}$ W m$^{-2}$\\
         Planet internal heat flux, $F_{\mathrm{int}}$ & 3500 W m$^{-2}$\\
         Optical absorption coefficient, $\kappa_{vis}$ & $2.4 \times 10^{-2}$ cm$^{2}$ g$^{-1}$ \\
         Infrared absorption coefficient, $\kappa_{IR}$ & $1 \times 10^{-2}$ cm $^{2}$ g$^{-1}$ \\
         Specific gas constant, $R$ & 3523 J kg$^{-1}$ K$^{-1} $\\
         Ratio of gas constant to heat capacity, $R/c_{p}$ & 0.286 \\
\enddata
\tablenotetext{a}{Assumed to be in a synchronous rotation state.}
\end{deluxetable}
Other commonalities of the models include the addition of ``sponge layers'' in the top three layers of the model. These layers apply extra drag near the top boundary of our model to reduce the buildup of artificial numerical noise. These layers have little effect on the atmospheric flow below them and can be found in many GCMs  \citep[see][for examples]{Mayne2014, Deitrick_2020THOR, Wang_2020}.

\subsection{Drag treatment in our three models}
The three models of WASP-76b from which we have chosen to generate emission spectra have different treatments of drag: Drag-free/0G, Uniform/$10^{4}$s, and Active/3G.  Here, we will summarize the main distinctions of these treatments. For further information about the detailed differences in circulation patterns that result from these drag treatments, see \citet{Beltz2022}. We review the main observable differences in Section \ref{sec:modelresults}.

\begin{itemize}

\item Drag-free/0G: This model serves as a base comparison to the other two models as there is no additional sources of drag aside from the sponge layers restricted to the top three layers and the numerical hyperdissipation that prevents the build up of noise on the smallest scales \citep[][]{Thrastarson2011}. Both of these artificial sources of drag also exist in the other models and are standard features of many GCMs. 

\item Uniform/$10^{4}$s: This model uses a simplified treatment of drag in which the same drag timescale (in this case, $10^{4}$ s) is applied globally at every level to remove momentum from the east-west and north-south components. When serving as a proxy for magnetic effects however, this method has the unfortunate implication of assuming a stronger surface magnetic field (or stronger coupling)  on the nightside of the planet than the dayside, contrary to what is expected physically \citep{Beltz2022}. Uniform drag is used in many GCMs and can also be used to model the effects of other large-scale atmospheric events like hydrodynamic shocks or turbulence \citep[][]{Li2010,PerezBecker2013}. 

\item Active Drag/3G: This treatment of drag is the more physically consistent than the uniform model and falls under the ``kinematic'' MHD umbrella. Instead of using a single global timescale, we locally calculate our active drag timescale based on the following expression from \citet{Perna2010magdrag}:
\begin{equation} \label{tdrag}
    \tau_{mag}(B,\rho,T, \phi) = \frac{4 \pi \rho \ \eta (\rho, T)}{B^{2} |sin(\phi) | }
\end{equation}
where $B$ is the chosen global magnetic field strength (in this case 3~G), $\phi$ is the latitude, $\rho$ is the density,  and the magnetic resistivity ($\eta$) is calculated in the same way as \citet{Menou_2012}: 
\begin{equation} \label{resistivity}
    \eta = 230 \sqrt{T} / x_{e} \textnormal{ cm$^{2}$ s$^{-1}$}
\end{equation} 
where the ionization fraction, $x_{e}$, is calculated from the Saha equation, taking into account the first ionization potential of all elements from hydrogen to nickel \citep[as in][]{RauscherMenou2013}. The derivation of this timescale assumes that the planet's interior generates a dipole field, aligned with the planetary rotation axis, that is significantly stronger than any induced magnetic field in the modeled atmosphere \citep{Perna2010magdrag}. As a result, the active drag is only applied in the east-west direction. 
This active drag timescale was first used in GCMs of the hot Jupiters HD~189733b and HD~209458b \citep{RauscherMenou2013}. In the ultrahot Jupiter regime, this timescale was used in \citet{Beltz2022} where we showed that the strong day-night temperature contrast of WASP-76b would result in the timescale \added{varying by many orders of magnitude. For example, at the $10^{-3}$ bar level, the timescale varies from $\sim 10^{18}$ s on the cold nightside to $\sim 10^{3}$ s on the dayside (see the right panel of Figure 1 in \citet[][]{Beltz2022} for the global distribution of timescales for our 3G model.) }\deleted{varying by nearly 15 orders of magnitude at a single pressure level.} This timescale is the most complex treatment of magnetic drag of the models presented here but still does not reach the complexity of implementing the full set of non-ideal MHD equations.
\end{itemize}

\subsection{Generating Emission Spectra from the  GCM}
In order to determine how different magnetic drag parameterizations in our GCM of the UHJ WASP-76b  manifest themselves in the high-resolution emission spectra we take our model output and post-process it with a detailed radiative transfer code. First, we  regrid the GCM's 65 vertical layers in pressure to 250 layers at constant altitude, by assuming  vertical hydrostatic equilibrium, consistent with  the GCM.\footnote{Because of the difference in scale heights the dayside vs the nightside, this results in some ``empty'' grid cells near the top of the nightside of the planet. By setting the temperature, winds, and opacity of these cells to 0, they are essentially ignored in our calculation} This is necessary in order to calculate line-of-sight columns and to increase the spatial resolution of the radiative transfer simulations. Based on resolution tests performed by \citet{Malsky2021}, this number of altitude layers should be sufficient to accurately calculate the emergent spectra. 

We perform the radiative transfer calculations for the post-processed spectra following the methods in \cite{Zhang2017}.  We implement the two-stream approximation for inhomogeneous multiple scattering atmospheres from \cite{Toon1989}. The 3D planet model is divided into 1-D line-of-sight columns (48 latitude and 96 longitude points) and the outgoing intensity from each column is calculated independently. For each column we calculate the total optical depth of each of the 250 layers. The model has the capabilities of simulating different cloud species \citep[e.g.][]{Harada2021}, but no clouds were included in this work. Therefore, 
the total optical depth at each wavelength is simply

\begin{equation}
\tau_{\lambda} = \int \kappa_{gas} \ dl
\end{equation}

\noindent where $\kappa_{gas}$ is the local gas opacity at each point along the line-of-sight path. The opacities are evaluated at their Doppler-shifted wavelengths, according to the line-of-sight velocity at each location:

\begin{eqnarray}
    v_{\mathrm{LOS}} = u \sin{\theta} + v \cos{\theta} \sin{\phi} - w \cos{\theta} \cos{\phi} \nonumber
    \\
    + \Omega(R_p+z) \sin{\theta} \cos{\phi}
\end{eqnarray}

\noindent where $\phi$ and $\theta$ correspond to the latitude and longitude, respectively, $\Omega$ is the planetary rotation rate, and $u$, $v$, and $w$ are the wind speeds in the east-west, north-south, and vertical directions, respectively.

\begin{figure}
    \centering
    \includegraphics[width=3.5in]{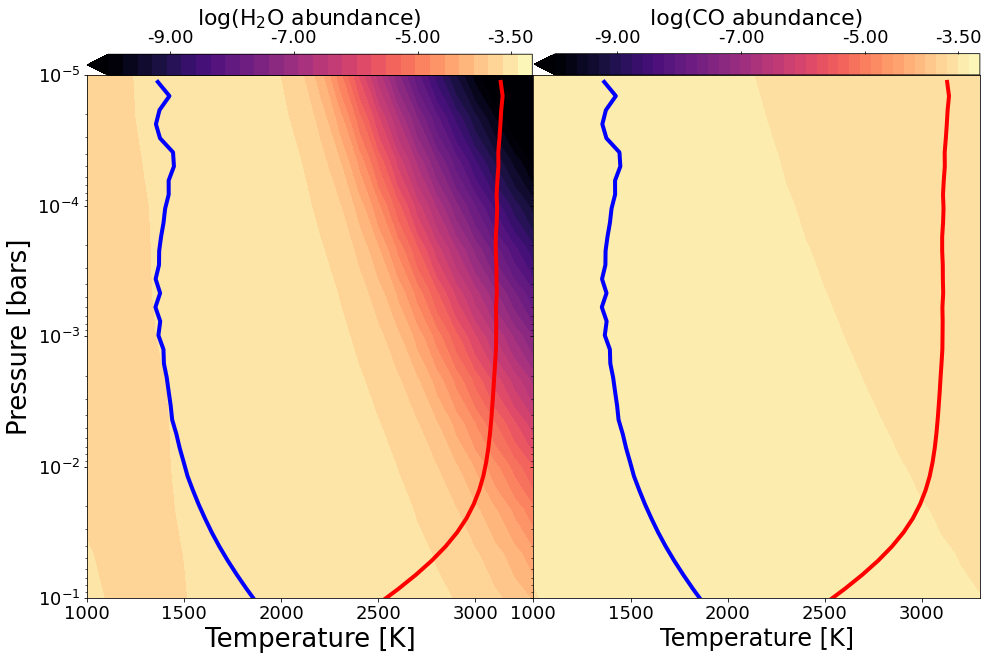}
    \caption{Abundance contours of water (\textit{left}) and CO (\textit{right}) as a function of temperature and pressure, used to generate emission spectra, with dayside (red) and nightside (blue) equatorial profiles overplotted. \deleted{ The dashed blue contours show abundance contours for values of $-3.8$ and $-3.4$.}  The strong temperature difference between the day and nightside profiles results in a large difference in water abundances. CO, on the other hand, varies less as a function of temperature and thus the CO abundances for the day and night side are similar.  } 
    \label{fig: WaterCOabun}
\end{figure}

In calculating our spectra, we assumed Local Chemical Equilibrium (LCE) and solar elemental abundances, which results in the amount of any particular molecular species varying as a function of temperature and pressure. In Figure \ref{fig: WaterCOabun}, we show the abundance contours for water and CO (the two main sources of opacity at the wavelength ranges presented in this work) as well as equatorial dayside and nightside  temperature-pressure profiles. \deleted{Also shown are lines of constant abundance for each species.}  Because of the strong temperature difference between the two hemispheres, we also expect a strong chemical gradient as well for some species. We can see from Figure \ref{fig: WaterCOabun} that the water abundances differ more strongly between the day and night side compared to CO. These differences in abundances will have implications for the line strengths in the corresponding emission spectra. 

For this work, we calculated spectra at two different wavelength ranges. The first of these spans from 1.14 $\mu m$ to 1.35 $\mu m$, which is inside the wavelength range covered by the upcoming WINERED instrument \citep{WINERED2016}. This wavelength range has features from H$_{2}$O, TiO, VO, K, and Na and most of the spectra and results we show here are from this wavelength range. The other wavelength range we explored was 2.3 to 2.35 $\mu m$, which contains opacities from CO, H$_{2}$O, and TiO and sits inside the range covered by IGRINS \citep[][]{Yuk2010,Park2014,Mace2016}, CRIRES, and CRIRES+ \citep[][]{CRIRES2004,CRIRES+2014}. We explore the differences between these two wavelength ranges in Section \ref{sec:wavelength_ranges}. Spectra were calculated from planetary snapshots at phases of every 11.25 degrees for the entire orbit. We use the GCM outputs for the cases of 0~G, 3~G, and the shortest uniform drag timescale ($10^{4}$ seconds) from \citet{Beltz2022}. We calculate the spectra for two different conditions: one in which the spectral lines are broadened and shifted due to winds and rotation (``Doppler on" spectra) and one without these effects (``Doppler off"). All spectra were calculated at a resolution of R=125,000.  


\section{Results} \label{sec: Results}
\subsection{3-D Atmospheric Structure} \label{sec:modelresults}
\begin{figure*}
    \centering
    \includegraphics[width=0.65\textwidth]{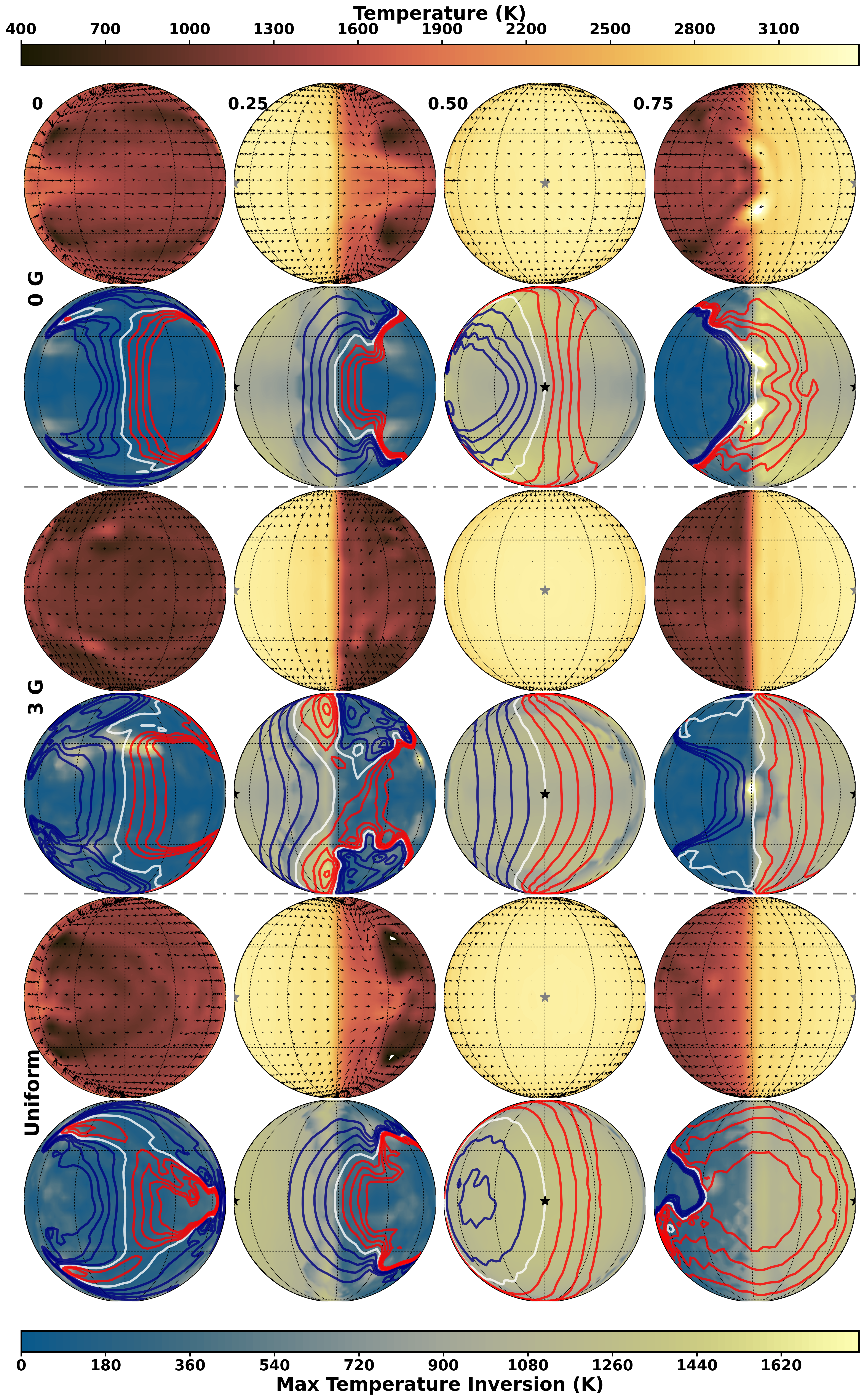}
    \caption{Orthographic projections for the 3 models presented in this paper (pairs of rows, top to bottom: 0~G, 3~G, and uniform) at 4 orbital phases (from left to right: 0, 0.25, 0.50, and 0.75; transit would be at 0). The first row of each model shows the temperature structure at $10^{-4}$ bars (within the region probed by spectral line cores) with wind directions plotted as arrows. The second row for each model shows the maximum vertical temperature inversion at each location. The blue and red contours show constant line of sight velocities in increments of  $\pm$2 km/s at the same pressure level as the temperature plot immediately above it. The differences in temperature structure and wind patterns will influence the high-resolution emission spectra generated from these models. }
    \label{fig:ortho projections }
\end{figure*}

Before diving into the spectra, we will briefly summarize the atmospheric structures of our models, as the temperature and wind structure will play a dominant role in shaping the resulting emission spectra. In Figure \ref{fig:ortho projections } we show the temperature and wind structures on the hemisphere facing an observer, at four phases throughout the planet's orbit and for each of the three models examined in this work. The temperature and wind maps (shown at $10^{-4}$ bars, around where spectral line cores form) help to visualize the differences in circulation patterns between the three models. The maps of where strong temperature inversions exist \citep[calculated as per][]{Harada2021} and the line of sight velocities (from winds and rotation) help to visualize which parts of the planet will emit spectra with emission or absorption features, and the net Doppler shift of those features.

Differences in temperature structures and atmospheric flow patterns between these models are discussed in detail in \citet{Beltz2022}. The main changes we saw once our active magnetic drag was turned on was an increase in the day-night temperature contrast, a decrease in the hotspot offset, and a change in the flow pattern. Under this active drag treatment, the dayside flow pattern switches from a day-to-night flow over most of the terminator to a flow that is channeled over the poles.  Some interesting features not discussed previously are found in  the line of sight velocities for our active magnetic drag model.  Especially at the phase of 0.25, one can see red-shifted regions near the terminator on the dayside blow up and over the pole, appearing as blue-shifted regions on the nightside.  In the uniform drag model we see that significant changes to the circulation over the terminator to the west of the substellar point results in some changes in the line-of-sight velocity patterns, particularly at phases of 0 and 0.75.

\subsection{Spectra as a Function of Orbital Phase}

\begin{figure}
    \centering
    \includegraphics[width=3.5in]{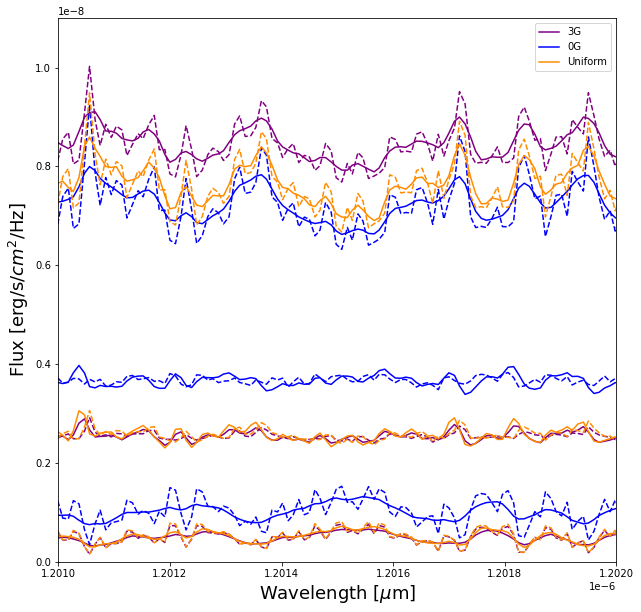}
    \caption{Spectra from our 3 models at an orbital phase of 0 degrees (bottom three spectra, when the nightside of the planet is in view), 0.25 (middle 3 spectra, when the day and night are both half in view), and 0.50 (top three spectra, when the dayside is in view). The solid lines show spectra that have been Doppler shifted by winds and rotation while the dashed lines are the spectra without this broadening. By comparing the relative fluxes of these spectra, we can see that the 3G model has the hottest dayside and coldest nightside of all the models shown here.     } 
    \label{fig: daynightallmodels}
\end{figure}

In Figure \ref{fig: daynightallmodels}, we compare spectra from our 3 different models at three orbital phases: 0 (where the observer only sees the nightside of the planet), 0.25 (first quadrature, where part of dayside and part of the nightside is in view), and 0.50 (where the viewer sees only the dayside of the planet). Due to the higher temperature of the dayside, the spectra corresponding to a phase of 0.50 have a much higher continuum flux than the nightside spectra. The continuum flux level for each model is set by the disk-integrated photospheric brightness temperature at each phase. We can see the largest day--night temperature difference in the 3~G model, followed by the uniform model, and the 0~G model, consistent with the simulated orbital phase curves from these models \citep[][]{Beltz2022}. Similarly, the 0~G model has a higher continuum flux at an orbital phase of 0.25, compared to the dragged models, since it more efficiently advects hot gas from the dayside toward the eastern terminator, which is the region in view at this phase. 

\begin{figure*}
    \centering
    \includegraphics[width=0.8\textwidth]{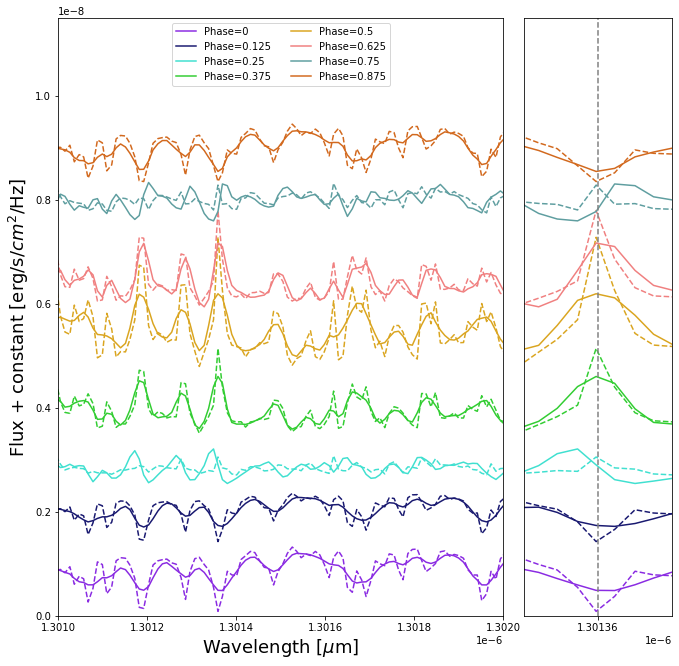}
    \caption{Emission spectra generated throughout the orbit of our 3G model planet. The solid lines show spectra that have been calculated including Doppler shifting from winds and rotation, while the dashed spectra do not include those effects. The bottom of the plot starts with the spectrum at a phase 0 (corresponding to transit, when the nightside of the planet faces us) and moving up the y-axis moves forward with time (including arbitrary offsets in flux to separate the spectra). \added{On the side panel on the right, a single feature is isolated with a grey vertical dashed line showing line center for the Doppler off case. } \deleted{When focusing on a single feature (e.g., the line near 1.302 $\mu$m), one can see} The feature begins as absorption at phase 0, due to the fact we are viewing entirely the nightside of the planet. Moving forward with time (increasing orbital phase), we can see the feature switch to emission as parts of the planet with temperature inversions come into view (with full dayside at a phase of 0.50) and then back to absorption again. Additionally, the line center is redshifted and blueshifted slightly throughout the orbit due to line-of-sight velocities from the planetary winds and rotation.      } 
    \label{fig: 3Gallphases}
\end{figure*}

In Figure \ref{fig: 3Gallphases}, we show spectra from our 3~G model at a variety of phases, both with and without the Doppler effects from winds and rotation. \added{\footnote{Throughout this work we show the summed component of winds and rotation. For this planet, the rotational velocity at the equator is $\sim 5.2$ km $s^{-1}$, which is slightly below the maximum speed of the equatorial jet from the 0~G model \citep[see Figure 4 in][]{Beltz2022}. Previous work \citep{Zhang2017,Beltz2021a} has studied the relative contributions of winds and rotation in high resolution emission spectra and found that the resulting broadening is more nuanced than just a sum of these velocities and is inherently 3D. Since they have comparable contributions to the line-of-sight velocities, it is important to include both.}} The spectra show an obvious switch between absorption features and emission features as different parts of the planet come into view. While the spectral lines begin in absorption when the nightside of the planet faces us, (at a phase of 0) by a phase of 0.375, enough of the dayside has come into view that the feature switches to emission, as there are temperature inversions found throughout the dayside in our model. Eventually, the features switch back to absorption at a phase of 0.875, as the enough of the nightside comes back into view.  Another interesting feature occurs at a phases of 0.25 and 0.75 where both absorption and emission features are present. Additionally, the emission features are blueshifted and the absorption features are redshifted at the orbital phase of 0.25, with the opposite direction in the shifts at 0.75.

\begin{figure*}
\begin{interactive}{animation}{othrogifwithpahse.mp4}
    \centering
    \includegraphics[width=5.25in]{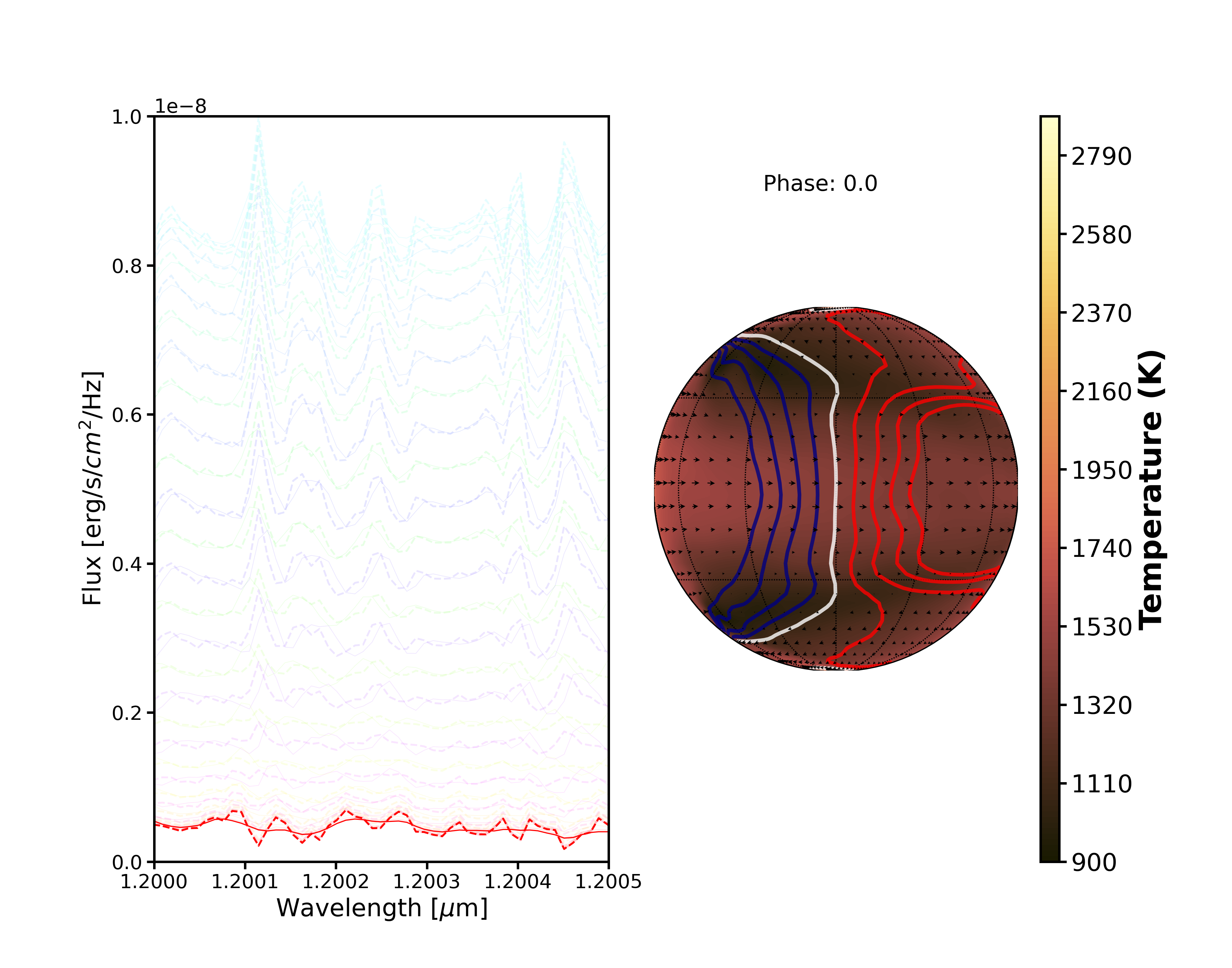}
\end{interactive}
    \caption{A comparison between the temperature and line-of-sight velocity structure visible on the planet (\textit{right}) and the disk-integrated emission spectrum (\textit{highlighted, left}) at each phase from our 3~G model. As the planet rotates and different regions come into view, the corresponding spectra will vary accordingly. Spectra with emission features are emitted from the hot dayside regions, where temperature inversions exist, while the cooler nightside regions emit spectra with absorption features. The line-of-sight velocities, due to winds and rotation, modulate the net Doppler shift of these different spectral components. This figure is available as an animation in the HTML version of the final article. The 10 second animation shows the planet rotating from a phase of 0 (shown in the static image) throughout its entire orbit. As the planet rotates, the temperature and wind fields change and the corresponding emission spectra is highlighted on the left. During the first half of the animation, the hotter dayside comes into view, increasing the continuum level of the emission spectra. At a phase near 0.5, the dayside of the planet is in view and the corresponding emission spectrum has the highest flux continuum level. Additionally, as the dayside of the planet comes into view the spectral features switch from absorption lines to emission lines. From phases 0.5 to 1.0, more of the nightside starts to come into view and the continuum level of the flux decreases and the spectral features switch back to absorption lines.   }
    \label{fig:gif}
\end{figure*}

We can understand this behavior by looking at Figure \ref{fig:ortho projections } and animated Figure \ref{fig:gif}, which highlights the spectra and corresponding temperature map at each phase. When looking at the line of sight velocities at a phase of 0.25 for example, we see both redshifted and blueshifted components near the limbs which results in the emission features possessing a different net shift than the absorption features coming from  the opposite limb. \added{This effect would only be present in spectra of sufficient spectral resolution as lower resolutions could result in flat, featureless spectra at these phases. We can roughly estimate the minimum spectral resolution needed to detect both the absorption and emission components by using $R \sim \frac{\lambda}{\Delta \lambda}$ where $\lambda$ is the line center of the non-Doppler shifted spectra and $\Delta \lambda$ is the difference between the peak of the emission feature and the trough of the absorption feature. When applied to the feature shown in Figure \ref{fig: 3Gallphases}, this roughly corresponds to a minimum resolution of $\sim 50,000$.  }Note that in the spectra calculated without Doppler shifts, the absorption and emission features almost balance out, resulting in largely featureless spectra. By taking Doppler shifts \added{at sufficient resolution }into account, we do not lose the spectral line information.  \added{The high-resolution emission spectra we present have inherently three-dimensional effects which would not be produced from a one-dimensional model. These effects are especially important and become more computationally complex when considering non-transiting planets \citep[as shown in][]{Malsky2021}. In non-transiting cases, even at phase of 0.5, part of the nightside hemisphere will be visible which influences the resulting high-resolution spectra. As more high-resolution observations are taken, it will be critical to take into account multi-dimensional effects.    }

\begin{figure}
    \centering
    \includegraphics[width=3.25in]{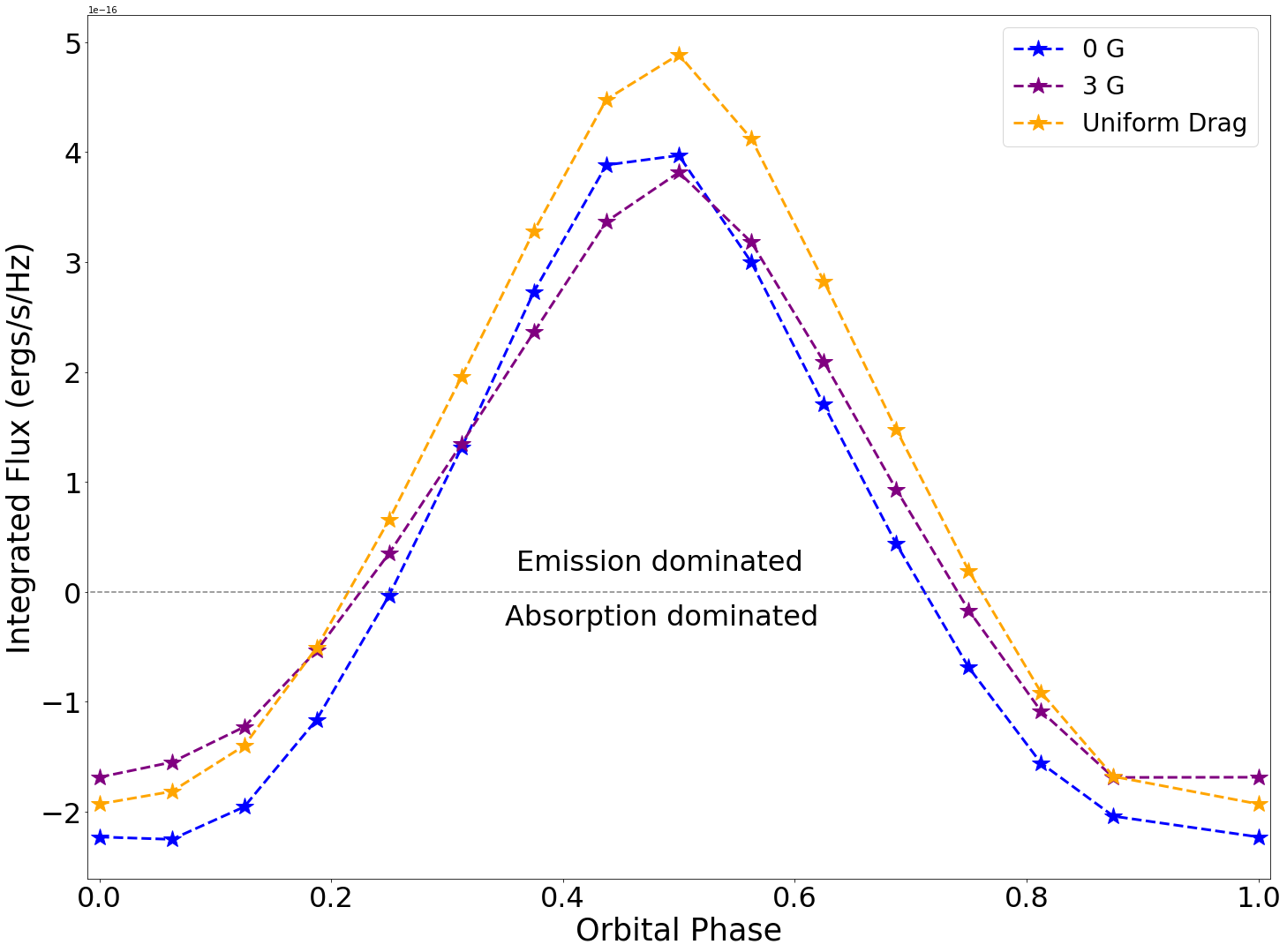}
    \caption{The integrated line flux for the three models in this paper, as a function of orbital phase.  From this plot, we see that the emission lines from the uniform drag models are consistently the strongest in the emission dominated regime and the 0~G model has the strongest absorption features. Another interesting aspect is that the model spectra switch between being dominated by emission vs absorption lines at different orbital phases.       } 
    \label{fig: intflux}
\end{figure}

While Figures \ref{fig: daynightallmodels} and \ref{fig: 3Gallphases} allow us to view by eye the change in absorption features to emission features as a function of phase, it is helpful to examine this trend quantitatively. By generating spectra with and without water, we can subtract one from the other to isolate the water features, both in emission (positive) and absorption (negative). Using numerical integration, the absorption and emission components are summed at each phase. Spectra without Doppler effects were used to generate this plot, so all differences are a result of differing temperature structures between the models.  In Figure \ref{fig: intflux}, we show the integrated line flux (positive for emission, negative for absorption) for water features (the most prominent source of opacity within the 1.14 $\mu$m to 1.35 $\mu$m wavelength range) from our 3 models. The shape of all three curves match basic predictions: emission peaks at an orbital phase of 0.50 and absorption peaks at a phase of 0. Interestingly, the uniform drag model consistently shows the strongest integrated emission while the 0~G has the strongest integrated absorption. Another detail to note from this figure is that the models do not transition from emission-dominated to absorption-dominated at the same phase. These differences in the intensity and orbital phase dependence of spectral features are a result of complexities in the models' corresponding temperature structures, which vary based on the strength and type of magnetic drag applied.  

\subsection{Cross-Correlation Results}

Another way to combine information from multiple lines within the planet spectra is to use cross-correlation to effectively integrate over all spectral lines, which is typically how signals are extracted from HRS observations. As described previously, two types of spectra are generated at each phase: Doppler on (which contains broadening and line shifting from winds and planetary rotation) and Doppler off (which contains none of this broadening). By cross-correlating the Doppler on version of the spectrum with the Doppler off version (in velocity space), the peak of the cross-correlation function (CCF) will tell us the net Doppler shift of the signal, while the width of the cross-correlation function quantifies the total amount of broadening in the spectrum \citep[see Figure 9 in][for example]{Beltz2021a}.

\begin{figure*}
    \centering
    \includegraphics[width=0.85\textwidth]{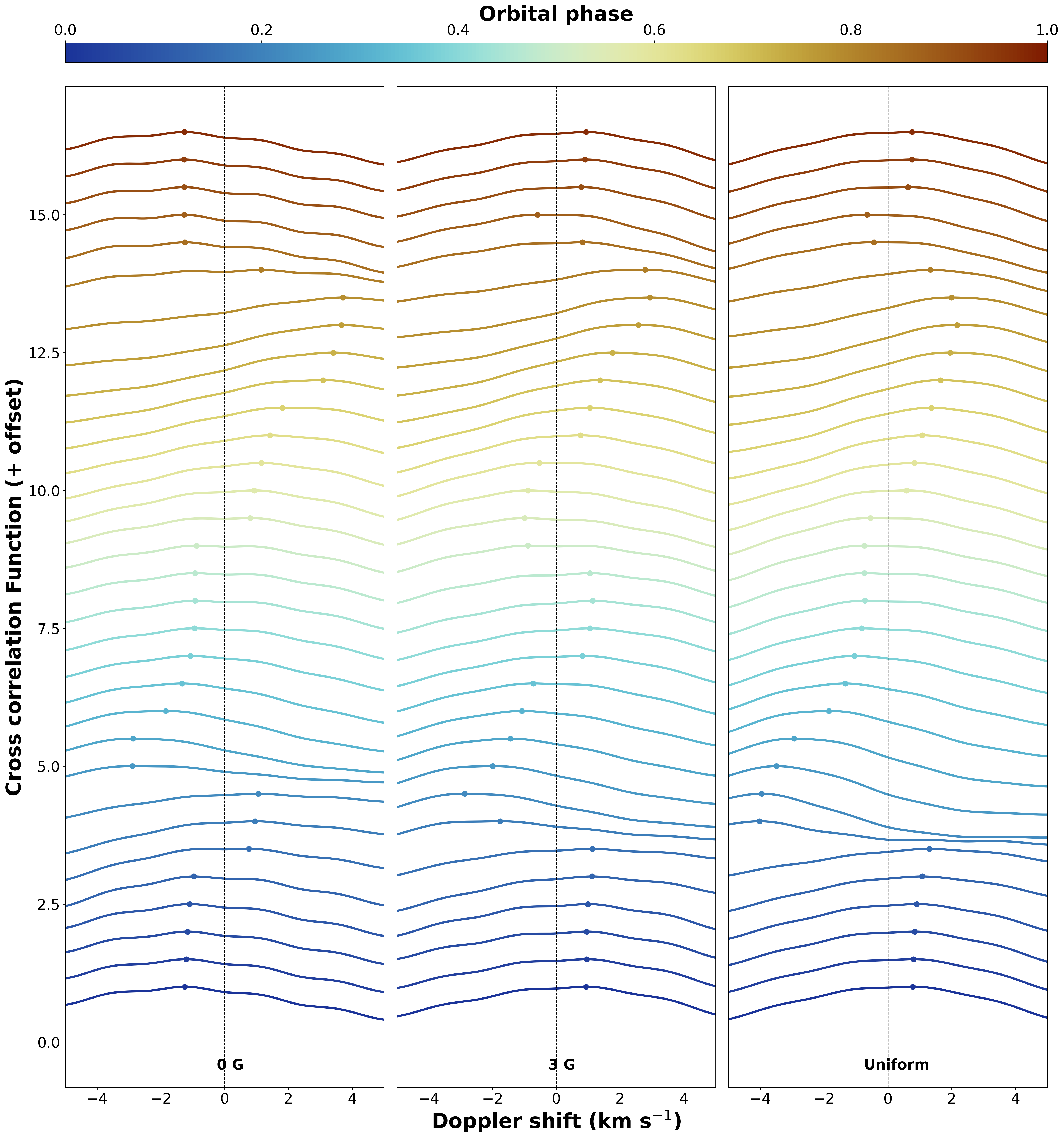}
    \caption{Cross-correlation functions of the spectra from our 3 different models throughout the planet's orbit. Each  Doppler-shifted spectrum was cross-correlated with its unbroadened counterpoint, with the peak cross correlation value denoted by the colored points. The peak of the cross correlation function quantifies the net shift of that particular spectrum. When different parts of the planet are in view, the net Doppler shift of the corresponding spectra will vary. The drag-free model differs in net Doppler shift from the active and uniform drag models  early and late in orbital phase, when much of the nightside is observable. Our active magnetic drag model shows a reversal in the net shift versus orbital phase trend around secondary eclipse (phase of 0.50) that is not seen in the drag-free and uniform drag model.    }
    \label{fig:cc}
\end{figure*}

Figure \ref{fig:cc} shows a series of cross-correlation functions from each of our 3 GCMs. 
At each phase, different parts of the planet's temperature and velocity fields are within the observable hemisphere, which influences the broadening and net shift of the spectra. \citet{Zhang2017} first calculated simulated high-resolution emission spectra from 3-D models of three hot Jupiters (without any added drag) and found some similarities in the patterns of net Doppler shift versus orbital phase between those planets (their Figure 7). We see that our Doppler shifts reach higher absolute values, likely due to the more extreme winds and faster rotation of our UHJ; our 0~G model is the most similar to their results for WASP-43b, the fastest rotator (and one of the more highly irradiated) of their set. Similarly to \citet{Zhang2017}, the net Doppler shift varies in sign and magnitude throughout orbit, as different parts of the planet rotate into view. However, the spectra presented here have their peak redshifts slightly later in phase ($\sim 0.75$) compared to the ones from \citet{Zhang2017} which peak in redshift closer to a phase near $0.61$.
 
These net Doppler shifts are informed by the temperature structure of the planet, which we see in Figure \ref{fig:ortho projections } and animated Figure \ref{fig:gif}. For example, looking at a phase of 0, we see that the the drag-free model has a warmer, brighter region on the left because the equatorial jet has advected that gas around from the dayside, which aligns with the blueshifted velocity structure, resulting in a net blueshifted CCF value at this phase.  In contrast, the temperature pattern on the nightsides of the dragged models is more complex, with cooler regions somewhat associated with blueshifted structure, ultimately resulting in net redshifts.   At phases where much of the dayside is in view ($\sim$0.25-0.75), the spectra from drag-free and uniform drag models show similar behavior in their net Doppler shifts, with a slow trend of increasing net shift with orbital phase (i.e., the spectra move from blue- to red-shifted). Our active drag model, however, displays more complex behavior, where this overall trend is disrupted by a superimposed net redshift to blueshift pattern right around 0.50. At phases where much of the nightside is in view ($< 0.25$ and$ > 0.75$) the drag-free model shows opposite net shifts than the 3~G and uniform model. This is a significant result: The net Doppler shift in nightside spectra may be indicative of the presence of magnetic drag (if redshifted) or not (if blueshifted).  A thorough analysis of velocity precision for HRS emission spectra has yet to be done, but theoretical work on simulated CRIRES data in \citet{Brogi2019} constrain water and CO's corresponding velocity with errors as small as $\pm 0.93$ km s$^{-1}$ and  $\pm 0.26$ km s$^{-1}$, respectively. At this level of precision, the differences between the Doppler shifts of our models would be feasible to detect.   However, we do caution that there may be some wavelength-dependence to these predictions (see Section \ref{sec:wavelength_ranges}). This is complementary to evidence for magnetic drag that could appear in net Doppler shift versus line strength trends in transmission spectra \citep{Kempton2012}.

\begin{figure*}
    \centering
    \includegraphics[width=6.5in]{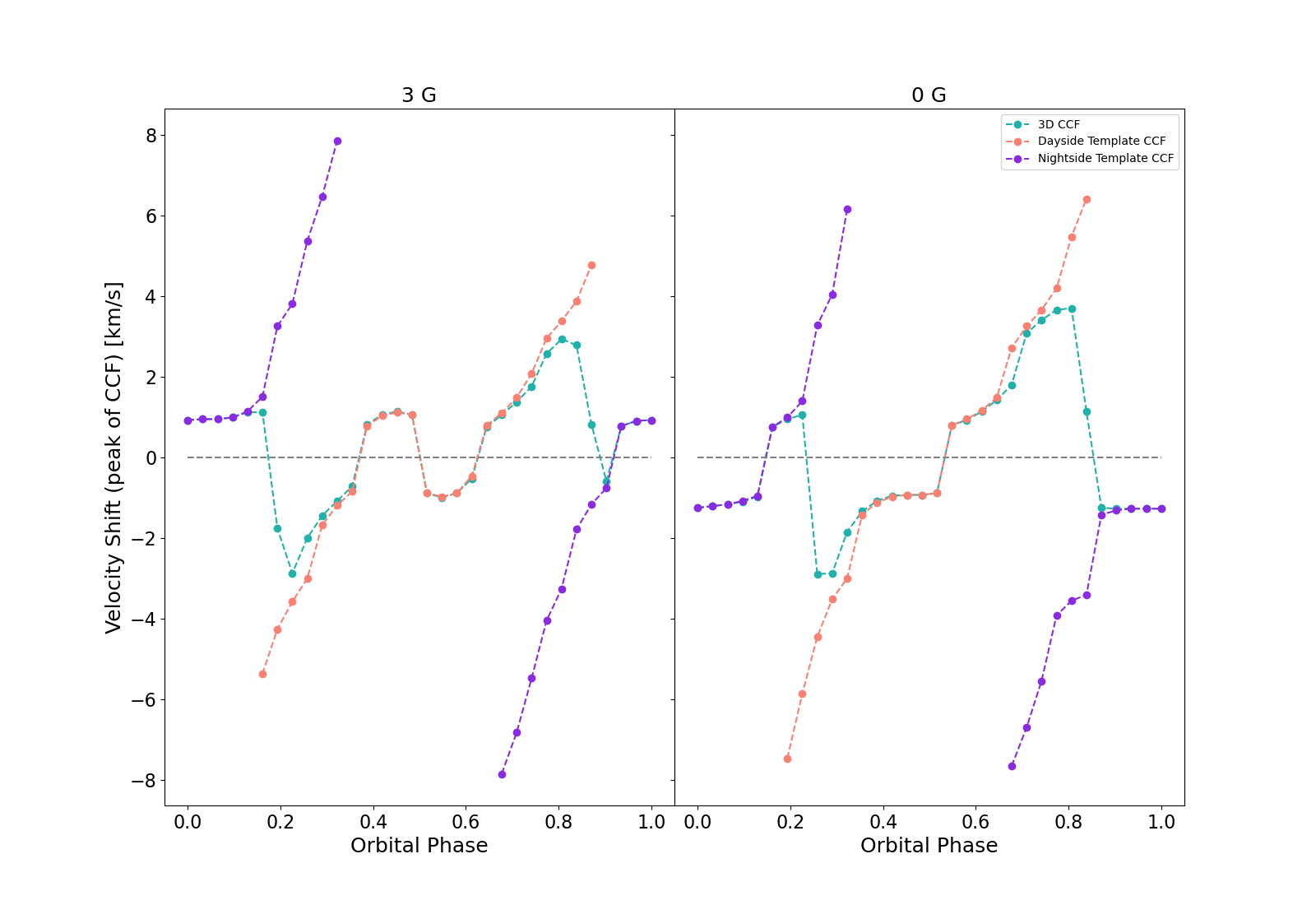}
    \caption{Effective retrieved net Doppler shift when using spectra generated as a function of phase (\added{teal} curve) compared to using only a dayside template (orange) or a nightside template (\added{purple}) against our 3~G model (left) and our 0~G model (right). The blue curve can be thought of as the ``true'' velocity shifts. While the dayside and nightside models match near their respective orbital phases, once they begin to be used at different phases, their accuracy decreases. At some points in the orbit, such as phases of 0.25 and 0.75, the day and nightside templates will return net shifts of opposite signs. To avoid introducing these errors, spectral templates should be generated as close to the true phase of observation as possible.    }
    \label{fig:3Dvs1Dcc}
\end{figure*}

Figure \ref{fig:3Dvs1Dcc} highlights the danger of ignoring the multidimensionality of planets, as expressed in their emission spectra. The \added{teal} curve, which reproduces the peak location of the CCF at each phase, as shown in Figure \ref{fig:cc},  shows the ``true'' function of peak velocity shift as a function of orbital phase. The orange curve was generated by cross-correlating a single dayside template spectrum---calculated from the 3D model at a phase of 0.50---with the spectra from our 3D model at each phase. The \added{purple} curve similarly does this with a single nightside spectrum---corresponding to a phase of 0. At phases near the template spectrum's origin phase, the peak velocity shifts match well with the blue curve. However, the farther away from the origin phase the single template is used, the larger the discrepancy from the true peak velocity. Most notably, at particular phases such as 0.25 and 0.75, the day and nightside templates return net velocity shifts of different signs and at very high net velocity, as these template spectra are just picking out the most red- or blue-shifted component of the observed hemisphere, as the day- or nightside region rotates in or out of view (see Figure~\ref{fig:ortho projections }).

This figure serves as a warning to observers to ensure the template spectra they use in their data analysis are generated at phases close to the observations. Longer observations over large ranges of orbital phase should use multiple templates for best results and to avoid misleading retrieved net Doppler shifts. While here we find different velocity components due to spatial variation in the planet's thermal structure, a similar effect resulting from spatial variation in chemical abundances may have been seen in \citet{Cont2021}, who detected two atmospheric species with significant differences in associated velocities in the UHJ WASP-33b. They interpret this large velocity offset to be a result of depletion of TiO (but not Fe) near the substellar point while both species exist on the terminator of the planet. Thus, their cross correlation technique may have detected TiO only near the terminator with large line-of-sight velocity components, which led to a higher velocity offset than the more evenly distributed Fe. 

\subsection{Cross-Correlation at Different Wavelength Ranges} \label{sec:wavelength_ranges}
\begin{figure*}
    \centering
    \includegraphics[width=6in]{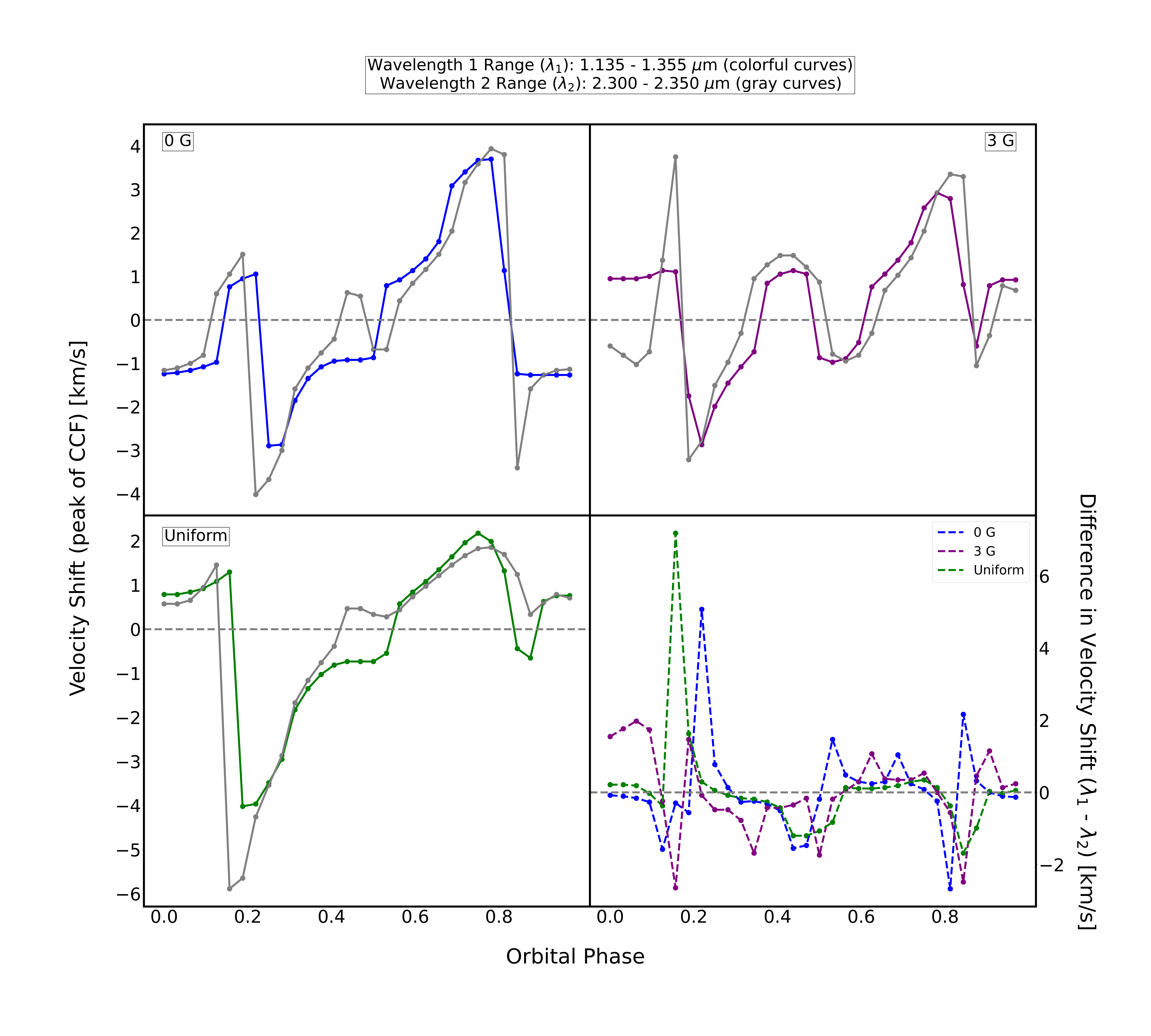}
    \caption{Net Doppler shifts versus orbital phase for emission spectra over the two wavelength ranges modeled. The colored lines show the Doppler shifts for wavelength range one, where water is the dominant absorber. The grey lines correspond to the shifts of wavelength range two, where CO is the dominant absorber. The different wavelength ranges probe different pressures in the atmosphere and, due to the temperature dependence of the water abundance, the disparity between the wavelength ranges should be spatially dependent. As such, the net Doppler shifts differ slightly. The bottom right panel shows this difference in net Doppler shift as a function of phase.   }
    \label{fig:ccdiffwl}
\end{figure*}

As discussed in Section \ref{sec: Methods}, we calculated spectra at two different wavelength ranges: 1.135--1.355 $\mu$m (wavelength 1, $\lambda_{1}$), used in all analyses up until this point, and 2.3--2.35$\mu$m (wavelength 2, $\lambda_{2}$). In Figure \ref{fig:ccdiffwl}, we show the Doppler shifts of the three different models at the two different wavelength ranges. Notably, the velocity shifts differ by some amount at each orbital phase tested, in some cases altering the trends identified above. This is a result of the different dominant atmospheric species in the two wavelengths ranges. In $\lambda_{1}$, the dominant absorbing species is water, while in $\lambda_2$ it is CO. These two species may be probing different atmospheric pressure levels, which can have different corresponding wind speeds and will result in different net Doppler shifts. In addition, the water abundance varies strongly as a function of temperature, while the CO abundance does not (see Figure \ref{fig: WaterCOabun}). Since the water opacity relative to CO opacity will therefore be location dependent, this means that the difference between the pressures probed by each wavelength range should also vary with the spatial location around the planet. This is a good demonstration that the 3D structure of the planet's atmosphere influences its emission spectrum in subtle and complex ways.

\section{Discussion} \label{sec:Discussion}
Important caveats to our GCM are already discussed in great detail in \citet{Beltz2022}. We will briefly mention these points here, and invite the reader to examine that work for a more detailed explanation. 

Because we have chosen to focus on the effects of magnetic drag in our modeling and post-processing, we are unable to include all the physical processes expected to be present on UHJs. Potentially the most consequential of these is molecular hydrogen dissociation and recombination. This would reduce one of the main effects we see from our active magnetic drag, in that it would result in a reduction in the day--night temperature contrast \citep{Bell_2018,Tan_2019}.
We also chose not to include clouds in these models, but work from \citet{Roman_2021} suggests that if present, cloud coverage would be very minimal and confined to the nightside upper latitudes, due to the high temperatures on the planet. \added{Inclusion of these clouds could alter the nightside temperature structure, potentially warming and cooling at different atmospheric heights. Additionally, depending on the height and opacity of the cloud deck, the strength of absorption features coming from this region of the planet could be muted. The strength of these effects would vary as a function of phase but would likely be most influential near quadrature phases (0.25 and 0.75) where emission features and absorption features are comparable in magnitude and potentially lead to these spectra to become more emission-dominated.  }

An important caveat to keep in mind when interpreting these results is that our magnetic drag prescription is a physically-motivated approximation of the non-ideal MHD equations. Our work assumes the toroidal component of the magnetic field, that is, magnetic field lines in the latitudinal direction, is negligible. Instead, based on our dipole assumption, our magnetic field lines are solely in the poloidal (longitudinal) direction. Models devoted to solving MHD equations, like the ones presented in \citet{Rogers_2014b} and \citet{Hindle2021hotspotreversals}, treat both of these magnetic field components, but have their own limitations, such as pre-calculated local conductivities (that do not self-consistently update as the simulation runs), using a simplified treatment for heating and cooling due to radiative fluxes, and the inclusion of explicit viscosity (resulting in significantly slower wind speeds). Our models and post-processed spectra presented here straddle the bridge between GCMs using simplified uniform drag and dedicated non-ideal MHD models. Our work highlights the ongoing need for model-to-model comparison while still presenting testable observables to gauge the validity of our approximations. Specifically, the results from Figures \ref{fig:cc}, \ref{fig:3Dvs1Dcc}, and \ref{fig:ccdiffwl}, are examples of the type of analysis that, if compared to observational data, can test our magnetic drag scheme's ability to approximate the physical environment of UHJs.  

One of the crucial assumptions in our drag treatment is that any magnetic field induced in the atmosphere is much smaller than the global magnetic field, a relationship that holds when the magnetic Reynolds number $R_{m} < 1 $. As we show in Figure 1 of \citep{Beltz2022}, this assumption holds true for the vast majority of the atmosphere except for some parts of the dayside near the low pressure boundary of our model. In order to more physically consistently describe the magnetic effects in this region requires full non-ideal magnetohydrodynamics (MHD). Some 3D non-ideal MHD models exist for (ultra) hot Jupiter atmospheres \citep{Rogers_2014b,Rogers2017McElwain,Rogers2017,Hindle2021hotspotreversals}
and show that variability in the circulation pattern (including westward flow near the substellar point) and the strength of the induced magnetic field may occur in the high magnetic Reynolds number regime. 

Since the double-gray radiative transfer in our GCM is agnostic to the specifics of the atmospheric compositions, it was necessary to make assumptions regarding the abundances of species in the atmosphere for our calculation of emission spectra. For simplicity's sake we assumed solar metallicity with abundances set by local chemical equilibrium. Measuring bulk metallicity for an UHJ is difficult, but measurements of water composition \citep{Kreigberg2014} and CO \citep[][]{Line2021} as well as results from interior models \citep[][]{Thorngren2016,Cridland2019,Thorngren2019} suggest our choice of solar abundances is reasonable. Work on this planet from \citet{Deibert2021, Landman2021, Wardenier2021} all assume solar metallicity for their analysis.  \citet{May2021} suggest that a lower bulk metallicity could explain the low internal heat flux that is more consistent with the observed phasecurve. On the other hand, \citet{Taberno2021} retrieve a slightly higher than solar metallicity for the host star, which \citet{Cassayas2021} use as justification to also test super-solar composition models. With this uncertainty in mind, more work is necessary to truly understand the metallicity of this planet. As pertains to our results, increasing the metallicity could result in moving the photosphere to slightly lower pressure in our GCM, perhaps with minor changes to the circulation pattern, and would lead to stronger spectral features in the emission spectra. These changes would not likely alter the main results and conclusions presented here.  

Another important aspect to consider is that in our calculations we have assumed local thermal equilibrium (LTE). This assumption is true for a majority of the atmosphere modeled, except for perhaps near the top boundary of our model. \citet{Deibert2021} found their NLTE model to be a slightly better fit than their LTE model of WASP-76b, but both models significantly underestimated the line depths. Its important to note however that their work used transmission spectra, which probes higher up in the atmosphere than emission and therefore we would expect less NLTE effects in the emission spectra presented here. Other work on NLTE effects in UHJs have focused on the hottest known exoplanet to date, KELT-9b. For pressures greater than $\sim 10^{-6} $ bar (which encapsulates all of the pressures probed in our GCM) LTE and NLTE models were nearly identical in predicting hydrogen level populations. In \citet{Fossati2021}, they use NLTE as a framework for pressures less than $10^{-4}$ bars when modeling temperature-pressure profiles. For pressures greater than $10^{-4}$ bars, they note that the retrieved temperature profiles are very similar. Based on these results, we expect that for the pressures modeled in our GCM, LTE is a valid approximation.

\added{It is tempting to extrapolate the trends in net Doppler shifts for the different models presented here to uniquely identify evidence of magnetic effects or estimates of magnetic field strengths. However, even if we take the assumption that the planet's magnetic field is a dipole aligned with its rotation axis to be completely true, these are Doppler shifts for one planet at one wavelength range. Additional physics not included in our model, such as clouds, hydrogen dissociation, or more complex magnetic feedback mechanisms could alter the strength of the predicted Doppler shift. Future work exploring a variety of field strengths and sweeping over a range of planetary parameters would be necessary to draw more quantitative conclusions about minimum field strengths. We can note, however, that our 3G model is the closest match to the phase curve published in \citet{May2021}. One feasible potential observational test from this work would be the case where if the trend in net Doppler shift versus orbital phase can be measured to sufficient precision, we could differentiate between whether the drag can be approximated with a uniform timescale or is better matched by our more complex treatment. Future observational and theoretical work is therefore needed to determine how robust these predictions are.}

\section{Conclusion} \label{sec:Conclusion}
UHJs are planets with strong temperature and compositional gradients due to their extremely short orbital periods and thus are excellent examples of the critical need for 3D models to understand the planet's emission spectra fully. In this work, we explore these 3D effects and illuminate how different forms of drag manifest in high resolution emission spectra. Our main points are as follows:
\begin{itemize}
    \item The strong temperature and chemical gradient on WASP-76b results in high-resolution emission spectra that vary strongly as a function of phase, showing at times absorption features, emission features, or both. These are inherently 3D features, in particular as the emission and absorption components may have different net Doppler shifts due to their emergence from regions of the planet with different line-of-sight velocities.
    \item The thermally ionized atmosphere of an UHJ should experience a magnetic drag  acting on its winds; this influences the flow pattern in the upper atmosphere and the Doppler shifts in the planet's emission spectra. We predict that the presence of this drag should result in a slight net redshift in nightside emission spectra, compared to the slight net blueshift seen in the drag-free model (both shifts $\sim$1 km/s). In addition, the complex form of magnetic drag may be apparent, over a simpler uniform-timescale drag, in the trend of net Doppler shift versus phase when the dayside of the planet is in view (showing a reversal in behavior around secondary eclipse). We caution that these predictions are dependent on the wavelength of observation.
    \item Using a single atmospheric template for cross-correlation at multiple phases may be poor matches to the planet, as the spectral features inherently change with viewing geometry. This can bias the recovered net Doppler shifts, especially at phases where parts of both the day and nightside are visible. In the most egregious cases, the amplitude of the calculated net Doppler shift can misleadingly exceed several km/s, when the inverted or non-inverted region of the planet (producing emission or absorption features, respectively) only appears on the edge of the planet disk, where the line-of-sight velocities from winds and rotation are maximized. 
\end{itemize}
The extreme spatial variations in ultrahot Jupiter atmospheres, exacerbated by the complication of magnetic drag, influence their high-resolution emission spectra. While these signatures can be subtle and highly dependent on the detailed 3D structure of the atmosphere, they may in turn offer us an avenue for empirically constraining those detailed physical properties and processes, possibly with current instruments but likely within the reach of future measurements with Extremely Large Telescopes.

\section{Acknowledgments}
This work was generously supported by a grant from the Heising-Simons Foundation. Many of the calculations in this paper made use of the Great Lakes High Performance Computing Cluster maintained by the University of Michigan. The authors also thank the reviewer for their helpful comments which improved the quality of this work.

\bibliographystyle{aasjournal}
\bibliography{bib.bib}

\end{document}